\documentclass[conference,twocolumn]{IEEEtran}
\usepackage{multirow}

\usepackage{graphicx}

\usepackage{hyperref}
\usepackage{url} 
\usepackage{doi}
\usepackage{siunitx}
\usepackage{dblfloatfix}
\usepackage{here}
\usepackage{listings}
\usepackage{etoolbox}
\usepackage{tikz}
\usepackage{textcomp}

\IEEEoverridecommandlockouts

\newcommand\copyrighttext{%
  \footnotesize \textcopyright 
  2017 IEEE. Personal use of this material is permitted. 
  Permission from IEEE must be obtained for all other uses, in any current or future media, 
  including reprinting/republishing this material for advertising or promotional purposes, 
  creating new collective works, for resale or redistribution to servers or lists, 
  or reuse of any copyrighted component of this work in other works.}%
\newcommand\copyrightnotice{%
\begin{tikzpicture}[remember picture,overlay]%
\node[anchor=south,yshift=10pt] at (current page.south) {\fbox{\parbox{\dimexpr\textwidth-\fboxsep-\fboxrule\relax}{\copyrighttext}}};
 \end{tikzpicture}%
}
\begin{document}

%
\title{Practical Implementation of Lattice QCD Simulation \\
on Intel Xeon Phi Knights Landing}

\author{
\IEEEauthorblockN{Issaku Kanamori
\thanks{Submitted to LHAM'17 ``5th International Workshop on Legacy HPC Application Migration'' in
CANDAR'17 ``The Fifth International Symposium on Computing and
Networking'' and to appear in the proceedings.}%
}
\IEEEauthorblockA{Department of Physics,\\
Hiroshima University\\
Higashi-hiroshima 739-8526, Japan\\
Email: kanamori@hiroshima-u.ac.jp}
\and
\IEEEauthorblockN{Hideo Matsufuru}
\IEEEauthorblockA{Computing Research Center,\\
High Energy Accelerator Research Organization (KEK)\\
Oho 1-1, Tsukuba 305-0801, Japan\\
Email: hideo.matsufuru@kek.jp}
}

\maketitle

\pagestyle{plain}
\copyrightnotice

\begin{abstract}
We investigate implementation of lattice Quantum Chromodynamics (QCD)
code on the Intel Xeon Phi Knights Landing (KNL).
The most time consuming part of the numerical simulations of
lattice QCD is a solver of linear equation for a large sparse matrix
that represents the strong interaction among quarks.
To establish widely applicable prescriptions, we examine rather
general methods for the SIMD architecture of KNL, such as using
intrinsics and manual prefetching, to the matrix multiplication and
iterative solver algorithms.
Based on the performance measured on the Oakforest-PACS system,
we discuss the performance tuning on KNL as well as the code design
for facilitating such tuning on SIMD architecture and massively
parallel machines.
\end{abstract}

\section{Introduction}
\label{sec:Introduction}

Quantum Chromodynamics (QCD) is the fundamental theory of
the strong interaction among quarks.
Despite its essential roles in the elementary particle physics,
the lack of analytical method to treat its large coupling at
low energy (equivalently at long distance) prevents us from solving
QCD in general.
The lattice QCD, that is formulated on a 4-dimensional Euclidean
lattice, enables us to tackle such problems by numerical simulations
\cite{textbook}.
Quanti ing with the path integral formalism, the theory resembles
a statistical mechanical system to which the Monte Carlo methods
apply.
Typically the most time consuming part of the lattice QCD
simulations is solving a linear equation for a large sparse fermion
matrix that represents the interaction among quarks.
As the lattice size becomes large necessarily to precision
calculation, the numerical cost grows rapidly.
The lattice QCD simulations have been a typical problem
in high performance computing and conducted development of
supercomputers such as QCDPAX \cite{QCDPAX} and QCDOC \cite{QCDOC}.

There are two distinct trends in high performance machines.
One is accelerators, represented by GPUs, which is composed of
many cores of $O(1000)$.
The Intel Xeon Phi architecture is a variant of the other type,
massively parallel clusters, while it has potential to work
as an accelerator.
Knights Landing (KNL), the second generation of Xeon Phi series,
reinforced usability as a massive parallel machine.
Its performance is assured by the SIMD architecture.
Elaborated assignments of vector registers and exploiting
SIMD instructions are essential to achieve desired performance.

In this paper, we port a lattice QCD simulation code to
Intel Xeon Phi Knights Landing.
The aim of this work is not a state-of-the-art tuning on KNL,
but to establish prescriptions that are applicable to wide
part of the application with practical performance.
While a hot spot of the QCD simulation tends to concentrate in
small part of programs, there are plenty amount of code
for measuring various physical quantities that have been accumulated
in legacy codes.
One frequently needs to accelerate such a code on a new architecture.
Thus it is important to establish simple procedures to achieve
acceptable performance as well as implication to future development of code.
For this reason, we restrict ourselves in rather general methods:
change of data layout, application of Intel AVX-512 intrinsics
and prefetching.
As a testbed of our analysis, we choose two types of fermion
matrices that are widely used together with iterative linear
equation solver algorithms.

This paper is organized as follows.
The next section briefly introduce the linear equation system
in lattice QCD simulations with fermion matrices employed
in this work.
Features of KNL architecture are summarized in
Section~\ref{sec:KNL}.
Section~\ref{sec:implementation} describes the details of
our implementation.
In the following sections, we measure the performance of
individual fermion matrices and the whole linear equation solvers.
The last section discusses implication of our results.

\section{Lattice QCD simulation}
\label{sec:lattice_QCD}

\subsection{Structure of lattice QCD simulation}

For the formulation of lattice QCD and the principle of the numerical
simulation, there are many textbooks and reviews \cite{textbook}.
Thus we concentrate on the linear equation system for the fermion
matrix, to which high computational cost is required.

The lattice QCD theory consists of fermion (quark) fields
and a gauge (gluon) field.
The latter mediates interaction among quarks and are represented
by `link variables', $U_\mu(x)\in SU(3)$, where
$x=(x_1, x_2, x_3, x_4)$ stands for a lattice site and
$\mu$=1--4 is the spacetime direction.
In numerical simulations the lattice size is finite,
$x_\mu=1,2,\dots, L_\mu$.
The fermion field is represented as a complex vector on
lattice sites,
which carries 3 components of `color' and 4 components of `spinor',
thus in total 12,
degrees of freedom on each site.
The dynamics of fermion is governed by a functional 
$S_F=\sum_{x,y} \psi^\dag(x) D[U]^{-1}(x,y) \psi(y)$,
where $D[U]$ is a fermion matrix.
A Monte Carlo algorithm is applied to generate an ensemble
of the gauge field $\{U_\mu(x) \}$, that requires to solve
a linear equation $x = D^{-1} \psi$ many times.

\subsection{Fermion matrix}

There is a variety of the fermion operator $D[U]$, since its
requirement is to coincide with that of QCD in the continuum limit,
the lattice spacing $a\rightarrow 0$.
Each formulation has advantages and disadvantages.
As a common feature, the matrix is sparse because of the locality
of the interaction.
In this paper, we examine the following two types of fermion matrices.

\subsubsection{Wilson fermion matrix}

\begin{figure}[!t]
\centering
\includegraphics[width=4.1cm]{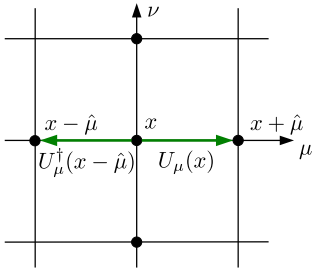}
\vspace{-1mm}
\caption{
The schematic feature of the Wilson fermion matrix.}
\label{fig:Wilson_fermion_matrix}
\end{figure}

The first one called the Wilson fermion matrix has the form
\vspace{-7mm}

{\small
\begin{eqnarray}
 D_W(x,y)\!\! &=& \!\!(m_0 + 4) \delta_{x,y} - \frac{1}{2} \sum_{\mu=1}^4
           \big[ (1-\gamma_\mu) U_\mu(x) \delta_{x+\hat{\mu},y}
\nonumber \\
& &
\hspace{16mm} 
        + (1+\gamma_\mu) U_\mu^\dag(x-\hat{\mu}) \delta_{x-\hat{\mu},y}
         \big] ,
\label{eq:Wilson_fermion_operator}
\vspace{-1mm}
\end{eqnarray}
\vspace{-5mm}
}

\noindent
where $x$, $y$ are lattice sites,
$\hat{\mu}$ the unit vector along $\mu$-th axis,
and $m_0$ the quark mass.
Fig.~\ref{fig:Wilson_fermion_matrix} indicates how the interaction
to the neighboring sites are involved in the matrix.
As mentioned above, the link variable $U_\mu(x)$ is 
a $3\times 3$ complex matrix acting on color
and $\gamma_\mu$ is a $4\times 4$
matrix acting on the spinor degrees of freedom.
Thus 
$D_W$ is a complex matrix of the rank $4 N_c L_x L_y L_z L_t$.
It is standard to impose the periodic or antiperiodic boundary
conditions.

\subsubsection{Domain-wall fermion matrix}

The second fermion matrix we treat is called the domain-wall
fermion.
This matrix $D_{DW}$ is defined by extending the spacetime to
a 5-dimensional lattice.
The structure of $D_{DW}$ in the 5th direction reads
\vspace{-2mm}

{\small
\begin{equation}
  D_{DW} = \left(
 \begin{array}{ccccc}
    D_W+1     &   -P_-     &     0       &\cdots &  m P_+     \\
  - P_+       &    D_W+1   &   - P_-     &       &   0        \\
   0          &   -P_+     &     D_W+1   &\ddots & \vdots     \\
   \vdots     &            & \ddots      &\ddots &   -P_-     \\
  m P_-       &   0        & \cdots      & - P_+ &    D_W+1   \\
\end{array}
\right) ,
\label{eq:domainwall_general}
\end{equation}
}
\vspace{-3mm}

\noindent
where $D_W(x,y)$ is the Wilson fermion matrix above
(with $m_0$ set to certain value), and $m$ instead represents
the quark mass.
$P_\pm$ are $4\times 4$ matrices
acting on the spinor components.
The size of the 5th direction, $N_s$, is also a parameter
of $D_{DW}$.
Eq.~(\ref{eq:domainwall_general}) means that $D_{DW}$ is a block
tridiagonal matrix including the boundary components
in the fifth direction.
Note that $U_\mu(x)$ is common in the 5th direction.
The domain-wall fermion is widely used because
of its good theoretical properties despite larger numerical
cost than the Wilson matrix.

There are several possible ways to implement $D_{DW}$.
The simplest way is to repeatedly use the implementation of $D_W$
to a set of 4-dimensional vectors.
Another way is to treat the structure in the 5th direction
as additional internal degrees of freedom, together with the
color and spinor ones.
The latter has an advantage in cache reuse.
We compare the both implementation and found that the latter 
achieves better performance, and thus concentrate on it in the following.

\subsubsection{Features of the fermion matrices}

\begin{table}[!t]
\renewcommand{\arraystretch}{1.3}
\caption{
Features of fermion matrices:
The number of floating point operation and the data transfer
between memory and processor per site.
For the domain-wall matrix, the 5th directional size is set as $N_s=8$.}
\label{comparison_of_fermion_matrix}
\centering
\begin{tabular}{ccccc}
\hline
 Fermion type & $N_{flop}$/site & data/site [B] (float) & Byte/Flop \\
\hline
 Wilson       & 1368  & 1536 B & 1.12  \\
 Domain-wall  & 11520 & 8256 B & 0.72  \\
\hline
\end{tabular}
\end{table}

Although these fermion matrices share the property of locality and
sparseness,
they have different features
in data size transferred between
the memory and the processor cores, number of arithmetic
operations, and data size of communication to other MPI process.
Table~\ref{comparison_of_fermion_matrix} summarizes the former
two values per site for single precision data.
For the domain-wall fermion matrix, these numbers
depend on the size of 5th direction, $N_s$.
Hereafter we adopt $N_s=8$ as a typical example.
As quoted in Table~\ref{comparison_of_fermion_matrix},
the domain-wall matrix tends to have smaller byte-per-flop
value, due to the independence of the link variable $U_\mu(x)$
in the 5th direction.

\subsection{Linear equation solver}

Since the fermion matrices are large and sparse,
iterative solvers based on the Krylov subspace method are
widely used.
For the Wilson fermion matrix, we employ the BiCGStab algorithm
for a nonhermitian matrix.
As for the domain-wall fermion, BiCGStab algorithm does not
work because its eigenvalues scatter also in the left side of
the imaginary axis.
We thus apply the conjugate gradient (CG) algorithm
to a hermitian positive matrix $D_{DW}^\dag D_{DW}$.

In practice a mixed precision solver is applicable.
In this case the single precision solver applied as the inner
solver determines the performance.
Therefore we measure the performance with the single precision.
While there are variety of improvement techniques for
a large-scale linear systems, such as a multi-grid or
domain-decomposition methods,
they are beyond the scope of this paper.

\section{Knights Landing architecture}
\label{sec:KNL}

The Knights Landing is the second generation of Intel Xeon Phi
architecture, whose details are found in \cite{KNLtextbook}.
Its maximal peak performance is 3 and 6 TFlops for double and single
precision, respectively.
It is composed of maximally 72 cores, in units of a
tile containing two cores.
Each tile has distributed L2 cache that is shared with 2 cores.
In addition to DDR4 with about 90 GB/s, MCDRAM of maximally 16 GB
accessible with 400 GB/s is available with one of three modes:
cache, flat, and hybrid.
Each core supports 4-way hyperthreading.
The SIMD instruction AVX-512 is available.
32 vector registers of 512-bit length are assigned to
8 double or 16 single precision numbers.

Our target machine is the Oakforest-PACS system hosted by
Joint Center for Advances High Performance Computing
(JCAHPC, University of Tokyo and University of Tsukuba)
\cite{Oakforest-PACS_website}.
The system is composed of 8208 nodes of Intel Xeon Phi
7250 (68 cores, 1.4 GHz) connected by full-bisection fat tree
network of the Intel Omni-Path interconnect.
It has 25 PFlops of peak performance in total, and started
public service in April 2017.

\section{Implementation}
\label{sec:implementation}

\subsection{Simulation code}

As the base code, we choose the Bridge++ code set
\cite{bridge_website,Ueda:2014zsa}, which is described
in C++ with the object-oriented design.
This code set allows us to replace fermion matrices
and solver algorithms independently.
Porting of Bridge++ to accelerators are performed in
\cite{Motoki:iccs2014}.
In the original Bridge++, hereafter called the
Bridge++ core library, the data is in double precision and
in fixed data layout.
Following the strategy employed in
\cite{Motoki:iccs2014}, 
we extend the vectors and matrices so as to enable any data
layout and changing the data type.
From the core library, the linear equation solver is offloaded
to the newly developed code.
This enables us to tune the hot spot on the specific architecture
with keeping the remaining part of Bridge++ available.
Following the implementation of Bridge++, we parallelized the
code with MPI and employ OpenMP for multi-threading.

\subsection{Related works}

Since the lattice QCD simulation, in particular the fermion
matrix multiplication, is considered a typical problem of
high performance computing, it is examined in a textbook
of KNL \cite{KNLtextbook}.
Its chapter 26 is devoted to performance tuning of
the Wilson-Dslash operator, corresponding to the Wilson matrix
above, using the QPhiX library \cite{QPhiX}.
On a single KNL with 68 cores at 1.4 GHz, the maximal performance
of the Wilson matrix multiplication achieved 272 and 587
GFlops for double and single precision, respectively.

As for the first generation of Xeon Phi, Knights Corner,
there are several works
\cite{Joo:2013,Li:2014kxa,Heybrock:2014iga,Arts:2015jia,
Boyle:2016lbp,Kobayashi:2016gog,Boku:2016dmw}.
Joo {\it et al.} \cite{Joo:2013} is the direct base of 
the QPhiX library on KNL.
Ref.~\cite{Boyle:2016lbp} developed a library named `Grid'
for the SIMD architecture, and has largely affected our work.
These works would be extended to the KNL as well.

\subsection{Implementation}

To fully exploit the SIMD architecture of KNL, rearrangement
of data is inevitably important.
For double and single precision data types,
512-bit register corresponds to 8 and 16 floating point
numbers, respectively, so we rearrange the date in these units.
We implement the code in C++ template classes and instantiate
them for double and float data types individually.
Since the vector data in lattice QCD are complex, there are
several possibilities for the ordering of real and imaginary parts.
Considering the number of SIMD registers and the number of
the degree of freedom on each site, we decide to place
the real and imaginary parts as consecutive data on the memory.
The color and spinor components are distributed to separate
registers.
For float type, data on eight sites
are packed into a single register and processed simultaneously.

\begin{figure}[!t]
\centering
\includegraphics[width=8.0cm]{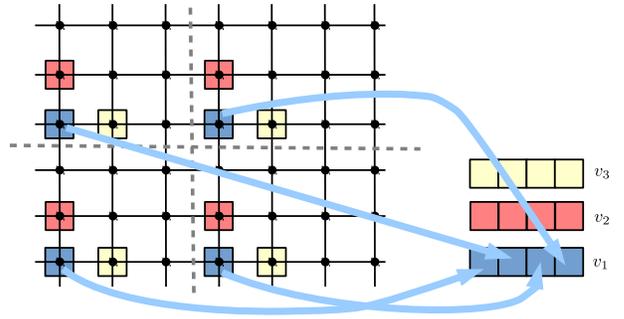}
\caption{
The site index ordering.  For float type complex variables, we use 3 dimensional
 analogues of this figure.}
\label{fig:GRID_site_ordering}
\end{figure}

There is flexibility how to fold lattice sites into the data
array.
We have tested several reasonable types of site ordering for
the Wilson fermion matrix and found similar performance.
Throughout this paper, we adopt one of them in which the 
implementation has been made most progress.
This site ordering is introduced in ~\cite{Boyle:2016lbp}.
Fig.~\ref{fig:GRID_site_ordering} displays how the index
of site is folded into SIMD-vector data.
This indexing does not require shuffling inside the
vector variables during the stencil calculation except for the boundary
sites in a local process, which makes the implementation easy.
It also allows us a flexible choice of local lattice volume.
In our implementation, local lattice sizes in $y$-, $z$-, and
$t$-directions must be even to pack 8 complex numbers into a single
SIMD-vector.
A possible disadvantage is a load imbalance between bulk and
boundary sites.
In most cases, however, this is hidden in the larger
imbalance due
to packing and unpacking of the data for MPI communication.

\subsection{Tuning procedure}

\subsubsection{Data alignment}

For a better performance with SIMD-vector data,
the data must be aligned by 64 bytes ({\it i.e.} 512 bits)
in AVX512 architecture.
To allocate the data, we use \texttt{std::vector} in the standard
C++ template library with providing an aligned allocator.
This allocation is done when only the master thread is
working.
The use of \texttt{std::vector} ensures that the data are
allocated on contiguous area of the memory.
For the KNL architecture, this may cause imbalance of
affinity between cores and memory and decrease of performance
in particular with large number of cores per process.
We nonetheless adopt this setup because of the least modification
of previously developed codes.

\subsubsection{Using Intrinsics}

The arithmetics for the SIMD-vector variables are implemented
with intrinsics.
For example, the following code is an implementation of
complex multiplication $a*b$ using AVX-512 intrinsics.

{ \small
\begin{lstlisting}
 // a, b are vector variables
 __m512 a_r=_mm512_moveldup_ps(a);
 __m512 a_i=_mm512_movehdup_ps(a);
 __m512 bt =_mm512_permute_ps(b,0xB1);
 a_i=_mm512_mul_ps(a_i, bt);
 __m512 c=_mm512_fmaddsub_ps(a_r,b,a_i);
\end{lstlisting}
}
\noindent
One can incorporate such codes with intrinsics by making use of
inline functions, C++ templates, or preprocessor macros.

We instead make use of \texttt{simd} directory in the `Grid'
library \cite{Boyle:2016lbp}, where a wrapper to the vector variable
(\texttt{\_\_m512}) and complex arithmetics are defined by using the intrinsics.
Using a template class defined there, operation $a*b$ can be written
such as \texttt{vComplexF c=a*b}.

\subsubsection{Prefetching}

We compare the manual prefetch and the automated prefetch by
compiler.
The most outer loop of the matrix is in the site index.
At each site, we accumulate 8 stencil contributions, from $+x$,
$-x$,...,$-t$ directions in order.
The prefetch to L1 and L2 cache is inserted
1 and 3 steps before the computation, respectively.
That is, before accumulating a $(+x)$-contribution, data
needed for $(-x)$-contribution is prefetched to the L1
cache and $(-y)$-contribution is to the L2 cache.
We use \texttt{\_mm\_prefetch} with
\texttt{\_MM\_HINT\_T0} and \texttt{\_MM\_HINT\_T1}
to generate the prefetch order.   
The following is a pseudo code to show the prefetch insertions:

{\small
\begin{lstlisting}
for(s=0; s<num_of_sites; s++){
#pragma noprefetch
  // +x
  prefetch_to_L1(-x);
  prefetch_to_L2(-y);
  accumlate_from(+x);
  // -x
  prefetch_to_L1(+y);
  prefetch_to_L2(+z);
  accumlate_from(-x);
  ...
}
\end{lstlisting}
}
\noindent
It is not straightforward to insert prefetch commands
appropriately.
One needs to tune the variables and the place to insert
referring to a profiler, {\it e.g.} Intel Vtune amplifier.
The performance may sensitive to the problem size,
choice of parameters such as the number of threads,
and so on.

\subsubsection{Thread task assignment}

Since the lattice extends over the machine nodes, 
the matrix and the reduction of vectors require communication
among nodes.
The function of matrix processes the following steps in order:
(1) Packing of the boundary data for communication,
(2-a) Doing communication,
(2-b) Operations of the bulk part, and
(3) Unpacking the boundary data.
(2-a) and (2-b) can be overlapped, and how efficient is this
is important for the total performance.

We restrict ourselves in the case that only the master thread
performs the communication, {\it i.e.} corresponding to
{\tt MPI\_THREAD\_FUNNELED}.
For the implementation of the steps (2-a) and (2-b) above,
there are two possibilities:
(i) arithmetic operational tasks are equally assigned to all
the available threads, and
(ii) the master thread concentrates the communication and
other threads bear the arithmetic tasks.
In this work, we adopt the case (ii).

\section{Performance of Wilson fermion matrix}

\subsection{Machine environment}

The performance is measured on the Oakforest-PACS
system.
We use the Intel compiler of version 17.0.4 with 
options {\tt -O3 -ipo -no-prec-div -xMIC-AVX512}.
On execution, we use job classes with the cache mode of MCDRAM.
According to the tuning-guide provided by JCAHPC,
we adopt the following setup.
To avoid OS jitter, the 0th and 1st cores on each KNL card
are not assigned for execution.
{KMP\_AFFINITY=compact} is set if more than 1 thread is assigned
to a core (unset for 1 thread/core).

\subsection{Wilson fermion matrix}

\subsubsection{Task assignment to threads}

We start with the Wilson fermion matrix.
We first compare the performance for combinations
of a number of cores per MPI process and a number of threads
per core.
For the former, 
(1) one core per process,
(2) two cores (one tile) per process, and
(3) 64 cores (whole KNL card) per process.
For the latter,
making use of the hyperthreading, the following three cases
are examined:
(a) one thread per core,
(b) two threads per core, and
(c) four threads per core.
The actual number of threads per MPI process is multiple of
numbers of cores and threads per core.

\begin{figure}[!t]
\centering
 \includegraphics[width=7.7cm]{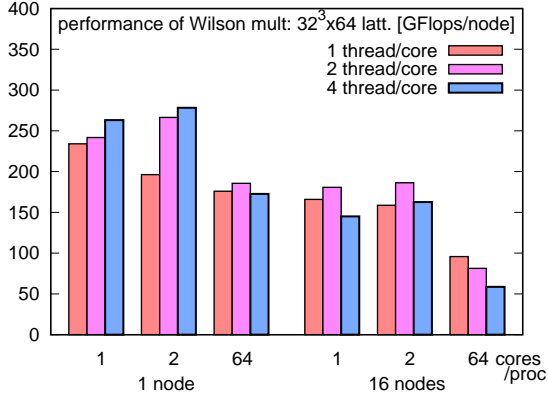}
\caption{
Comparison of numbers of threads per core.}
\label{fig:number_of_thread_per_core}
\end{figure}

We compare one, two, and four threads per core cases.
Fig.~\ref{fig:number_of_thread_per_core} shows the results with
1-node and 16-node, generated with the strong scaling on
$32^3\times 64$ lattice.
For the case~(3), copy of the boundaries is enforced so as to
compare with the other cases on the same footing.
For the case~(1) and (2), dependence on the number of threads
per core is not strong.
Including the cases of other numbers of nodes,
there is a tendency
of achieving 
the best performance with 2 threads per MPI process for
case~(1) and (2), and 1 thread per core for case~(3).

\subsubsection{Prefetching}

\begin{figure}[!t]
\centering
\includegraphics[width=7.7cm]{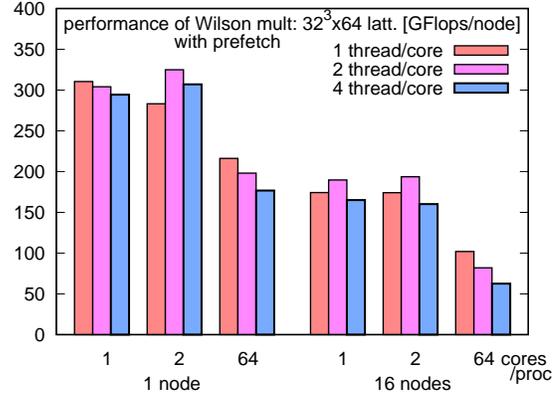}
\caption{
Effect of manual prefetching.
The performance of the Wilson matrix multiplication is measured on
a $32^3 \times 64$ lattice.}
\label{fig:Wilson_mult_effect_of_prefetch}
\end{figure}

Effect of manual prefetch against the automated prefetch by
compiler is displayed in
Fig.~\ref{fig:Wilson_mult_effect_of_prefetch}.
In the single node case, where no inter-node communication is needed,
the manual prefetch improves the performance by more than 20\%.
Contrary to a statement in \cite{KNLtextbook}, manual prefetching
is effective to our case.
Increasing the number of nodes, however, the effect is gradually
washed and becomes a few percent at 16 nodes as shown in
Fig.~\ref{fig:Wilson_mult_effect_of_prefetch}.
Since our target lattice sizes assumes more than $O(10)$ KNL nodes,
the advantage of manual prefetch is not manifest compared to
involved tuning effort.
For this reason, we do not apply it to the
linear algebraic functions and the domain-wall fermion matrix,
while the following measurement of the Wilson matrix is done
with the tuned code.

Here we summarize the output of Intel Vtune Amplifier for the
Wilson matrix multiplication with a single process of 64 threads.
Applying the manual prefetch, L2 cache hit rate increases
from 76.7\% to 99.1\%.
With the manual prefetch, the UTLB overhead and the page walk
are 0.0\% and 0.2\% of clockticks, respectively.
The metric SIMD Compute-to-L2 Access Ratio  reaches as large as 2,023.
For the whole BiCGStab solver examined below, however,
the L2 hit rate decreases to 84.4\%, as explained by the larger
byte-per-flop rate of the solver algorithm.

\subsubsection{Comparison to other codes}

Now we compare the performance of the Wilson matrix
multiplication to other codes under the condition
of a single process on a single KNL.
As quoted already, QPhiX library achieves 587 GFlops for
single precision \cite{KNLtextbook} on a $32^3\times 96$ lattice.
The GRID library \cite{Boyle:2016lbp} provides a benchmark
of the Wilson matrix that we run on the same environment as
this work.
On $32^3\times 64$ lattice, based on v0.7.0,
it gives the best performance with one thread/core and amounts
to 348.6 GFlops that is comparable to our result.
While our result is not as fast as QPhiX, it shows that
large fraction of performance can be achieved with widely
applicable techniques.
These values are reasonable considering the memory bandwidth
of MCDRAM and the byte-per-flop in
Table~\ref{comparison_of_fermion_matrix}, while far below
the peak performance.

For reference, we also measure the performance of the
original Bridge++, that is implemented in not SIMD-oriented
manner and only in the double precision.
The best performance is obtained with 4 threads/core and
results in 60.0 GFlops, which roughly corresponds to 120 GFlops
in single precision.
This indicates the impact of the SIMD-oriented tuning.

\subsubsection{Scaling property of matrix multiplication}

\begin{figure}[!t]
\centering
\includegraphics[width=8.0cm]{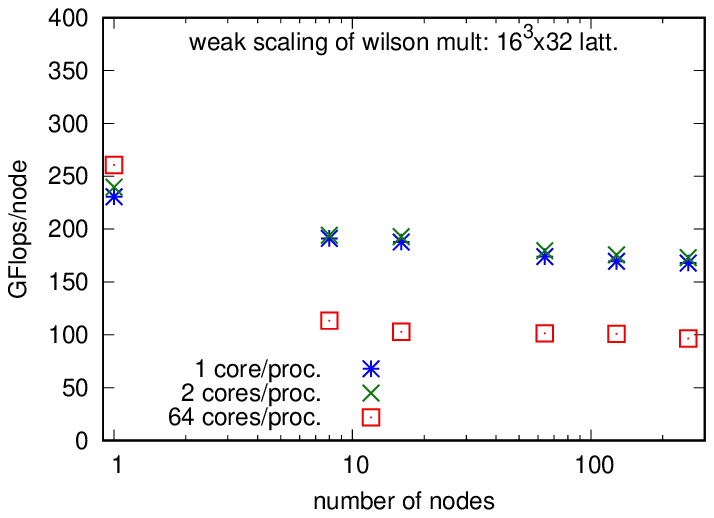}
\includegraphics[width=8.0cm]{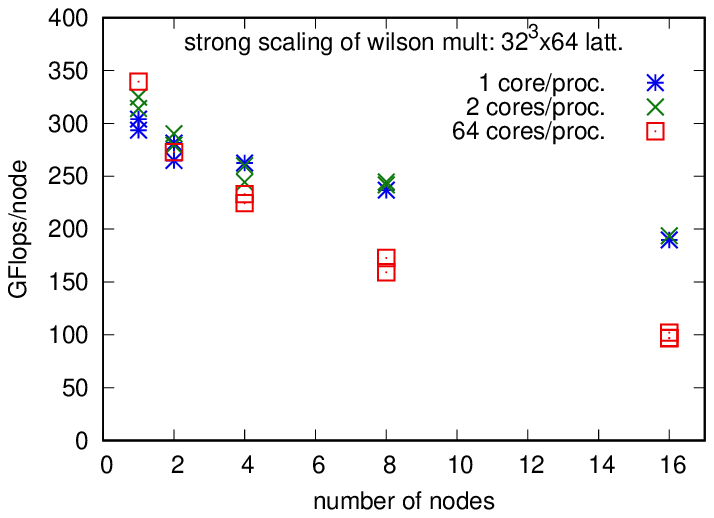}
\caption{
Scaling plots for the Wilson matrix multiplication.
Top: weak scaling with a $16^3 \times 32$ lattice in each node.
Bottom: strong scaling with a $32^3 \times 64$ lattice.
}
\label{fig:Wilson_mult_scaling}
\end{figure}

Now we observe the scaling properties of the Wilson matrix
multiplication.
In the following, we adopt 1, 2, and 2 threads/core for
64, 2, and 1 cores/process, respectively.
The top panel of Fig.~\ref{fig:Wilson_mult_scaling}
shows the weak scaling plot for the $16^3 \times 32$
lattice in each node.
In the measurements, we do not enforce the copy of the boundary data
unless it is really needed.
For reference, if it is enforced on a single node with 64 cores/process,
the performance becomes 100.4 and 176.0 GFlops on $16^3 \times 32$
and $32^3 \times 64$ lattices, respectively.
For the 64 cores/process on multiple nodes, the performance is
about the half of the other two cases.
This may be explained by that all the 64 cores in a node share
the whole memory of the node so that imbalance of accessibility
to the memory among the cores exists.
The cases of 1 or 2 cores/process achieve more than 170 GFlops/node
up to 256 nodes.

The bottom panel of
Fig.~\ref{fig:Wilson_mult_scaling}
shows the strong scaling on a $32^3 \times 64$
lattice.
In this case, as the number of nodes increases, the local lattice
volume inside each node decreases so that the communication
overhead becomes more and more significant.
Again the 64 cores/process case exhibits less performance than
other two cases.
For the strong scaling, the performance depends on how the
lattice is divided into sublattices.
We plot several cases at each number of nodes.
For the case of 1 and 2 cores/process, the performance at 16 nodes
is about 2/3 of that of single node.

\subsection{Performance of BiCGStab solver}

For the Wilson matrix, the BiCGStab solver works efficiently.
We compose the BiCGStab algorithm with BLAS-like library.
For the linear algebraic functions, we apply neither manual
prefetch nor additional compiler option for prefetch.
While the Wilson matrix part is improved by manual prefetch,
this effect is small because the performance is determined
by the linear algebraic functions.

\begin{figure}[!t]
\centering
\includegraphics[width=8.0cm]{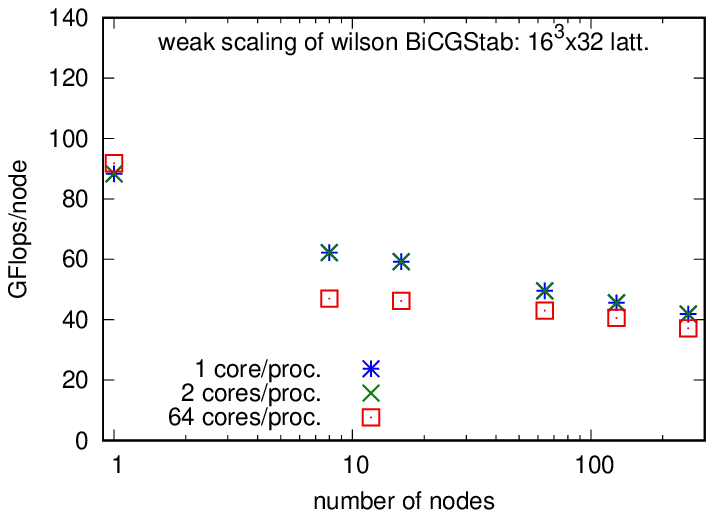}
\includegraphics[width=8.0cm]{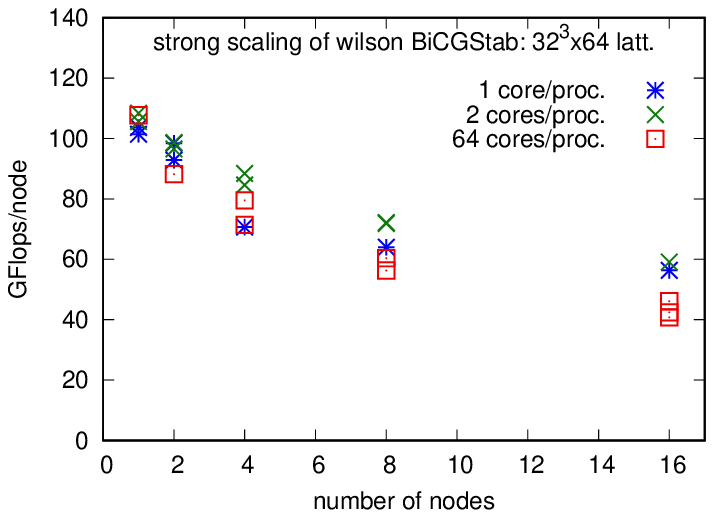}
\caption{
Scaling property of the BiCGStab solver with Wilson matrix.
Top: weak scaling plot for a $16^3 \times 32$ lattice in each node.
Bottom: strong scaling on a $32^3 \times 64$ lattice.
}
\label{fig:Wilson_solver_scaling}
\end{figure}

The top panel of Fig.~\ref{fig:Wilson_solver_scaling}
shows the weak scaling for the BiCGStab solver on
a $16^3 \times 32$ lattice in each node.
Because of larger byte-per-flop values of the linear algebraic
functions, the performance reduces to about 1/4 of the matrix
multiplication at 256 nodes.
The difference of 64 cores/process and other two cases also
decrease because of the linear algebraic functions.
The worse scaling as the number of node is caused by the
reduction operations.
The bottom panel of Fig.~\ref{fig:Wilson_solver_scaling}
shows the strong scaling plot of the solver on $32^3 \times 64$
lattice.
While the difference of 64 cores/process and other two cases
shrinks, smaller numbers of cores/process achieve better
performance for large number of nodes.
In total, these results indicate that small number of cores
per MPI process has advantage,
if the memory size and the local lattice volume allow.

\section{Performance of domain-wall fermion matrix}

\subsection{Performance of matrix multiplication}

For the domain-wall fermion matrix, the condition of
performance measurement is the same as the Wilson matrix
except for no manual prefetch is applied.
The domain-wall multiplication achieves better performance
than the Wilson matrix, as expected from smaller byte-per-flop
value due to the reuse of link variable $U_\mu(x)$,
as well as the larger number of arithmetic operations per
lattice site.

\begin{figure}[!t]
\centering
\includegraphics[width=8.0cm]{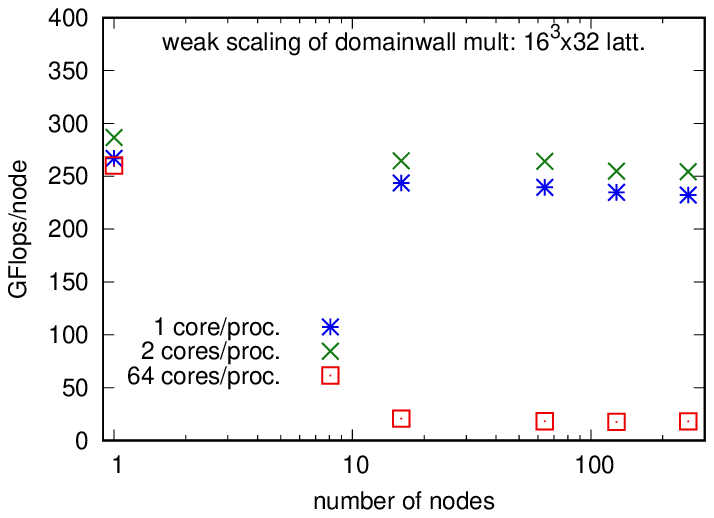}
\includegraphics[width=8.0cm]{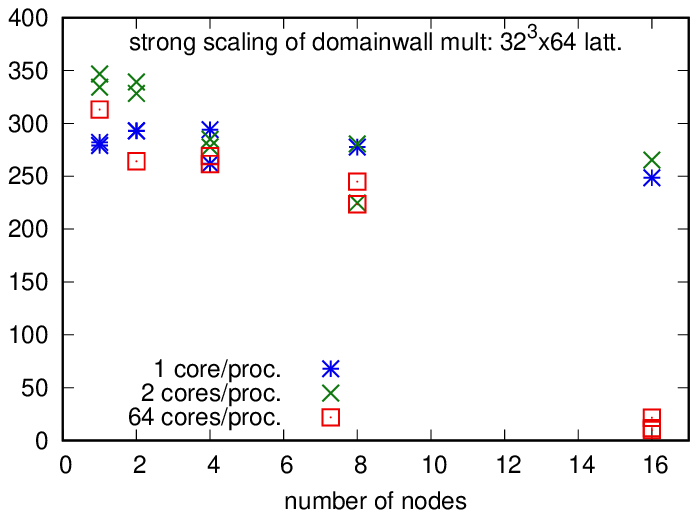}
\caption{
The weak and strong scaling plots of the domain-wall matrix multiplication.
For the weak scaling, $16^3 \times 32$ lattice per node.
For the strong scaling, $32^3 \times 64$ lattice.}
\label{fig:DW_mult_scaling}
\end{figure}

The top panel of Fig.~\ref{fig:DW_mult_scaling} displays
a weak scaling plot of the domain-wall matrix
multiplication for a $16^3 \times 32$ lattice in each node.
On a single node with 64 cores/process,
if the boundary data copy is enforced, the performance becomes
155.2 GFlops on a $16^3 \times 32$ lattice and 211.8 GFlops on
a $32^3 \times 64$ lattice for the weak and strong scaling,
respectively.
While the 1 and 2 cores/process cases exhibit good scaling
behavior, multi-node result of 64 cores/process is quite bad.
What is different from the Wilson matrix is larger size of
boundary data transferred at the communication.
How such reduction of performance occurs and how can be
avoided is now under investigation.
The strong scaling on a $32^3 \times 64$ lattice in the
bottom panel of Fig.~\ref{fig:DW_mult_scaling} shows
similar tendency as the Wilson matrix, except for the reduction
of the 64 cores/process data at 16 nodes.

For the 64 cores/process data, we observe large fluctuations
up to almost factor 10 in the elapsed time.
We have not understood why such fluctuations occur
and are investigating the cause.
For this reason, we do not include the data for 64 cores/process
in the measurement of the CG solver below.

\subsection{Performance of CG solver}

For the domain-wall fermion, we apply the CG algorithm to
the matrix $D^\dag D$.
Fig.~\ref{fig:DW_mult_scaling} displays the weak (top) and
strong (bottom) scaling plots for the CG solver.
Better scaling behaviors than those of the Wilson matrix
are explained by larger weight of 
the matrix multiplication in the algorithm and larger size
of vectors.
Even for the strong scaling plot, the performance
of 1 and 2 cores/process cases almost unchanged 
as increasing the number of nodes from 1 to 16.

\begin{figure}[!t]
\centering
\includegraphics[width=8.0cm]{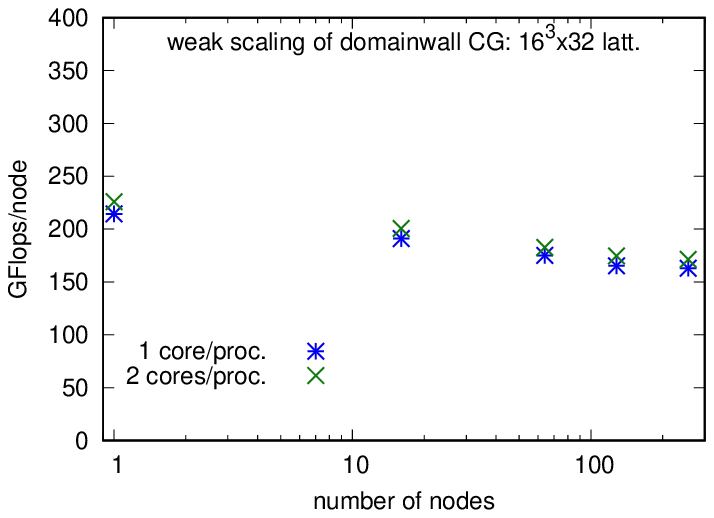}
\includegraphics[width=8.0cm]{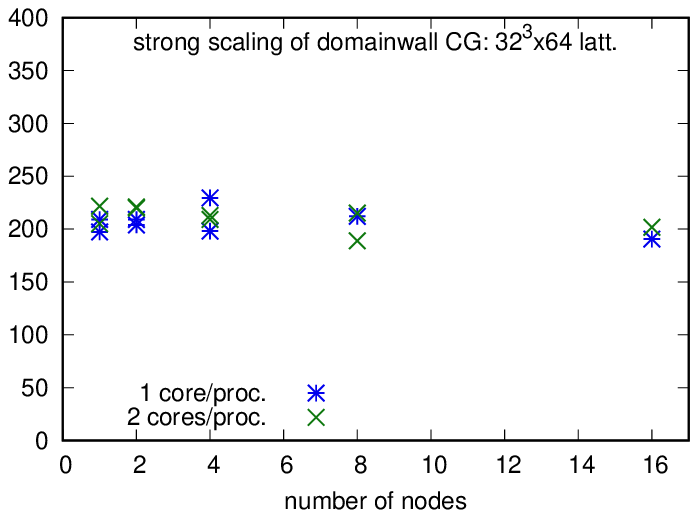}
\caption{
The weak and strong scaling plot for the CG solver with the
domain-wall fermion matrix.
For the weak scaling, $16^3 \times 32$ lattice per node.
For the strong scaling, $32^3 \times 64$ lattice.}
\label{fig:DW_solver_scaling}
\end{figure}

\section{Conclusion}

In this paper, we apply rather simple prescription to
make use of the SIMD architecture of the Intel Xeon
Phi KNL processor to a typical problem in lattice QCD
simulation.
Aiming at widely applying to existing codes,
we examine the rearrangement of data layout and AVX-512
intrinsics to arithmetic operations, and examined
the effect of manual prefetching.
The former two are inevitable to achieve acceptable
performance, compared to the original Bridge++ code.
The effect of manual prefetching is more restrictive.
It amounts to dedicated efforts only on single node
or small number of nodes.
Comparing the choices of numbers of cores per MPI
process, small numbers of cores per MPI process
have advantages as increasing number of nodes.

Our results indicate two efficient ways of using KNL.
On single KNL, multi-thread application without MPI
parallelization may works efficiently.
The manual prefetch is worth to try.
For a multi-node case, adopting small numbers of
cores per MPI process, like a massively parallel machine,
one can optimize the number of nodes against a given
problem size.
As for the code of application, it is essential to
employ design that enables flexible rearrangement of data
layout and incorporation of intrinsics.

\section*{Acknowledgment}

The authors would like to thank Peter Boyle, Guido Cossu,
Ken-Ichi Ishikawa, Daniel Richtmann, Tilo Wettig,
and the members of Bridge++ project for valuable discussion.
Numerical simulations were performed on Oakforest-PACS system
hosted by JCAHPC, with support of Interdisciplinary Computational
Science Program in CCS, University of Tsukuba.
This work is supported by JSPS KAKENHI (Grant Numbers
JP25400284, JP16H03988), and by
Priority Issue 9 to be tackled by Using Post K Computer,
and Joint Institute for Computational Fundamental Science (JICFuS).

\end{document}